\documentstyle[12pt,epsfig]{article}

\catcode`\@=11
\long\def\@makefntext#1{
\protect\noindent \hbox to 3.2pt {\hskip-.9pt  
$^{{\ninerm\@thefnmark}}$\hfil}#1\hfill}                

\def\@makefnmark{\hbox to 0pt{$^{\@thefnmark}$\hss}}  
        
\def\ps@myheadings{\let\@mkboth\@gobbletwo
\def\@oddhead{\hbox{}
\rightmark\hfil\ninerm\thepage}   
\def\@oddfoot{}\def\@evenhead{\ninerm\thepage\hfil
\leftmark\hbox{}}\def\@evenfoot{}
\def\sectionmark##1{}\def\subsectionmark##1{}}

\renewcommand{\thefootnote}{\fnsymbol{footnote}}

\newcounter{sectionc}\newcounter{subsectionc}\newcounter{subsubsectionc}
\renewcommand{\section}[1] {\vspace*{0.6cm}\addtocounter{sectionc}{1} 
\setcounter{subsectionc}{0}\setcounter{subsubsectionc}{0}\noindent 
        {\normalsize\bf\thesectionc. #1}\par\vspace*{0.4cm}}
\renewcommand{\subsection}[1] {\vspace*{0.6cm}\addtocounter{subsectionc}{1} 
        \setcounter{subsubsectionc}{0}\noindent 
        {\normalsize\it\thesectionc.\thesubsectionc. #1}\par\vspace*{0.4cm}}
\renewcommand{\subsubsection}[1]
{\vspace*{0.6cm}\addtocounter{subsubsectionc}{1}
        \noindent {\normalsize\rm\thesectionc.\thesubsectionc.\thesubsubsectionc. 
        #1}\par\vspace*{0.4cm}}

\newcounter{appendixc}
\newcounter{subappendixc}[appendixc]
\newcounter{subsubappendixc}[subappendixc]

\renewcommand{\appendix}[1] {\vspace*{0.6cm}
        \refstepcounter{appendixc}
        \setcounter{figure}{0}
        \setcounter{table}{0}
        \setcounter{equation}{0}
        \renewcommand{\thefigure}{\Alph{appendixc}.\arabic{figure}}
        \renewcommand{\thetable}{\Alph{appendixc}.\arabic{table}}
        \renewcommand{\theappendixc}{\Alph{appendixc}}
        \renewcommand{\theequation}{\Alph{appendixc}.\arabic{equation}}
        \noindent{\bf Appendix \theappendixc #1}\par\vspace*{0.4cm}}

\def\abstracts#1{{
        \centering{\begin{minipage}{12.2truecm}\footnotesize\baselineskip=12pt\noindent
        \centerline{\footnotesize ABSTRACT}\vspace*{0.3cm}
        \parindent=0pt #1
        \end{minipage}}\par}} 

\renewenvironment{thebibliography}[1]
        {\begin{list}{\arabic{enumi}.}
        {\usecounter{enumi}\setlength{\parsep}{0pt}
\setlength{\leftmargin 1.25cm}{\rightmargin 0pt}
         \setlength{\itemsep}{0pt} \settowidth
        {\labelwidth}{#1.}\sloppy}}{\end{list}}

\newcounter{itemlistc}
\newcounter{romanlistc}
\newcounter{alphlistc}
\newcounter{arabiclistc}

\newcommand{\fcaption}[1]{
        \refstepcounter{figure}
        \setbox\@tempboxa = \hbox{\footnotesize Fig.~\thefigure. #1}
        \ifdim \wd\@tempboxa > 6in
           {\begin{center}
        \parbox{6in}{\footnotesize\baselineskip=12pt Fig.~\thefigure. #1}
            \end{center}}
        \else
             {\begin{center}
             {\footnotesize Fig.~\thefigure. #1}
              \end{center}}
        \fi}

\newcommand{\tcaption}[1]{
        \refstepcounter{table}
        \setbox\@tempboxa = \hbox{\footnotesize Table~\thetable. #1}
        \ifdim \wd\@tempboxa > 6in
           {\begin{center}
        \parbox{6in}{\footnotesize\baselineskip=12pt Table~\thetable. #1}
            \end{center}}
        \else
             {\begin{center}
             {\footnotesize Table~\thetable. #1}
              \end{center}}
        \fi}

\def\@citex[#1]#2{\if@filesw\immediate\write\@auxout
        {\string\citation{#2}}\fi
\def\@citea{}\@cite{\@for\@citeb:=#2\do
        {\@citea\def\@citea{,}\@ifundefined
        {b@\@citeb}{{\bf ?}\@warning
        {Citation `\@citeb' on page \thepage \space undefined}}
        {\csname b@\@citeb\endcsname}}}{#1}}

\newif\if@cghi
\def\cite{\@cghitrue\@ifnextchar [{\@tempswatrue
        \@citex}{\@tempswafalse\@citex[]}}
\def\citelow{\@cghifalse\@ifnextchar [{\@tempswatrue
        \@citex}{\@tempswafalse\@citex[]}}
\def\@cite#1#2{{$\null^{#1}$\if@tempswa\typeout
        {IJCGA warning: optional citation argument 
        ignored: `#2'} \fi}}

 1
 1
 1

\font\ninerm=cmr9

\begin{document}

\vspace*{-2cm}
\centerline{\normalsize\bf THEORETICAL OVERVIEW OF JET PHOTOPRODUCTION AT HERA}

\vspace*{0.6cm}
\centerline{\footnotesize MICHAEL KLASEN}
\baselineskip=13pt
\centerline{\footnotesize\it Deutsches Elektronen-Synchrotron DESY, Theory
Group,}
\baselineskip=12pt
\centerline{\footnotesize\it Notkestrasse 85, D-22607 Hamburg, Germany}
\centerline{\footnotesize E-mail: michael.klasen@desy.de}

\vspace*{0.9cm}
\abstracts{We review the theoretical framework of jet photoproduction at HERA
discussing the conceptual ideas, phenomenological models, and higher order
perturbative calculations. Numerically, we study the physically interesting
distribution of transverse energy within the observed jet, the real and virtual
photon structure, and the proton structure at large $x$.}

\vspace*{0.5cm}
{\it To appear in Proc. of ``New Trends in HERA Physics'', Ringberg, May 1997.}

\normalsize\baselineskip=15pt
\setcounter{footnote}{0}
\renewcommand{\thefootnote}{\alph{footnote}}
\section{Theoretical Concepts}
In electron-proton scattering at HERA the dominant fraction of the scattering
events proceeds through photons with low virtuality $Q^2$.%
\footnote{From 1994-1997, HERA operated with positrons instead of electrons.
Since we will only be concerned with neutral current exchange, we use
the term ``electron'' for positrons as well.}
Experimentally the electron is anti-tagged and remains in the beam-pipe.
Theoretically the lepton tensor and phase space can then be factorized with
the Weizs\"acker-Williams approximation\cite{x1}, which was improved recently
through power-suppressed terms by Frixione et al.\cite{x2} The virtuality of
the photon is less than 4 GeV$^2$, and it retains a fraction $y \in [0.2;0.85]$
of the incident electron energy $E_e=27.5$ GeV. The proton energy is 820 GeV.

Jet production in photon-proton collisions was first calculated in leading
order (LO) of perturbative QCD by Owens\cite{x3}. The two-fold nature of
direct and resolved processes and their separation on a kinematical basis
were pointed out as a means to test the underlying partonic dynamics. It
is thus possible to study the distribution of partons in the initial photon
and proton and the spin of the exchanged particle.

Since then much effort has been spent on improving the tree-level
understanding. Phenomenological models have been implemented into the Monte
Carlo event generators PYTHIA\cite{x4}, HERWIG\cite{x5}, and PHOJET\cite{x6}
employing parton showers, fragmentation models, and multiple interactions.
Next-to-leading order (NLO) calculations for inclusive single-jet\cite{x7} and
dijet production\cite{x8} provide perturbative correction factors, reduce the
scheme and scale dependences, and allow for an implementation of jet
definitions. Recently the first NLO analysis of the transition from real to
virtual photoproduction of jets has been published\cite{x9}.

Although several parametrizations exist for the parton distributions in the
photon in NLO\cite{x10}, the gluon is still poorly constrained by the data
from $\gamma\gamma$ scattering at $e^+e^-$ colliders and needs complementary
measurements from $\gamma p$. The situation is better for the parton densities
in the proton\cite{x11}, where photoproduction data might improve the data
from deep inelastic and $p\overline{p}$ scattering at intermediate to large
values of $x$. Except where indicated we use the GS96 and CTEQ4M
parametrizations for the photon and proton, respectively.

\section{Phenomenological Models}
Perturbative QCD describes only the hard partonic scattering process and the
scale evolution of the hadronic structure functions. The link between colored
partons and real hadrons belongs to the non-perturbative domain and has to be
fitted to data or guessed from phenomenological models. These models are
implemented in Monte Carlo generators and compared in Table 1.
\begin{table}[h]
\tcaption{Properties of different Monte Carlo generators.}
\small
\begin{center}
\begin{tabular}{|c|c|c|c|}
\hline
 Monte Carlo & Parton        & Fragmen- & Multiple \\
 Generator   & Showers       & tation   & Interactions \\
\hline
\hline
 PYTHIA      & Initial+Final & String   & Hard (optional) \\
\hline
 HERWIG      & Initial+Final & Cluster  & Soft (optional) \\
\hline
 PHOJET      &         Final & String   & Soft and Hard   \\
\hline
\end{tabular}
\end{center}
\end{table}

As a first step, one can attach additional angularly ordered partons to the
hard process until the original parton reaches a maximum virtuality of
$Q_{\max}^2 < E_T^2$. In addition and deviating from the parton model,
intrinsic transverse momenta are allowed for the partons in the hadrons up to
$k_T < k_T^{\max}$ thereby introducing a second phenomenological parameter.

The second step consists in the fragmentation of the proliferated partons into
hadrons. The Lund string model confines the color field between quarks into a
color flux tube. The energy increases proportional to the distance between the
quarks until the string breaks up and new quark-antiquark pairs are created.
Hadrons are formed when the energy is too low for the string to break up
according to a fragmentation function with two free parameters. Gluons appear 
as excitations and produce kinks in the string. Alternatively the cluster
model starts with the splitting of gluons into $q\overline{q}$ pairs and the
subsequent formation of color neutral clusters. Heavy clusters cascade into
light clusters which then transform isotropically into hadrons.

Finally multiple interactions seem to be important at HERA to describe
transverse energy flow and cross sections in the direction of the proton
remnant at low $E_T$. They increase the multiplicity and energy flux and model
interactions between the photon and the proton remnant. Secondary interactions
are softer than the first scattering defined by $E_T$ and can be of partonic or
soft nature.

\section{Next-to-Leading Order Calculations}
The undesirable drawback of phenomenological models is the large number of free
parameters that have to be tuned to data. These are not present in
next-to-leading order calculations where one calculates virtual corrections
with internal particle loops and real corrections with soft and collinear
radiation one order higher in the strong coupling constant $\alpha_s$. The
singularities can then be controlled through dimensional regularization and
removed consistently through renormalization and factorization procedures.
This avoids unphysical cut-off parameters, reduces the scheme and scale
dependence, and allows for an implementation of jet definitions.

Real corrections are most easily calculated with the phase space slicing
method. After approximation of the invariants and partial fractioning of the
$2\rightarrow 3$ matrix elements, we factorized the Born process and integrated
the remaining singular kernels analytically up to an invariant mass cut-off
$y_{\rm cut}$. Aurenche et al.\cite{x8} employed a transverse energy cut-off
for the soft/collinear initial state, which does not appear in a related
method for the single-jet case, and a cone cut-off for the collinear final
state\footnote{See also the contribution by M.~Fontannaz in these
proceedings.}.
Harris and Owens integrated the soft and collinear regions separately\cite{x8}.
The real emission outside the cut-offs is integrated numerically. This removes
the dependence on the technical cut-off and introduces the experimental jet
definition.

The subtraction method relies on a point-by-point subtraction of singularities
in the numerical integration and has only been applied to the direct process by
B\"odeker\cite{x7}. The resolved process could, however, be adapted from an
existing program in $p\overline{p}$ scattering\cite{x12}.

\section{Jet Definition Uncertainties}
In hadronic collisions cluster algorithms of the JADE type combine particles
not only into the hard jets but also into the remnant jets. This is avoided if
a cone of size $R=1$ in azimuth-rapidity space is used to define a jet. The
Snowmass\cite{x13} accord
determines the jet axis from the $E_T$ weighted directions of all particles in
the cone. Unfortunately it contains a number of ambiguities which render a
matching between theory and experiment difficult\cite{x14}.

Contrary to fixed cone algorithms, iterative cone algorithms can merge
overlapping jets. This is not described by a NLO calculation with just three
final state partons. If two partons have a distance between $R$ and $2R$, they
can be counted as one or two jets, and one has to avoid double-counting. The
phenomenological parameter $R_{\rm sep}$ defining the distance of two partons
can be used to model the narrower jets found by iterative cone algorithms.

\begin{figure}[t]
 \begin{center}
  {\unitlength1cm
  \begin{picture}(14,8.5)
   \epsfig{file=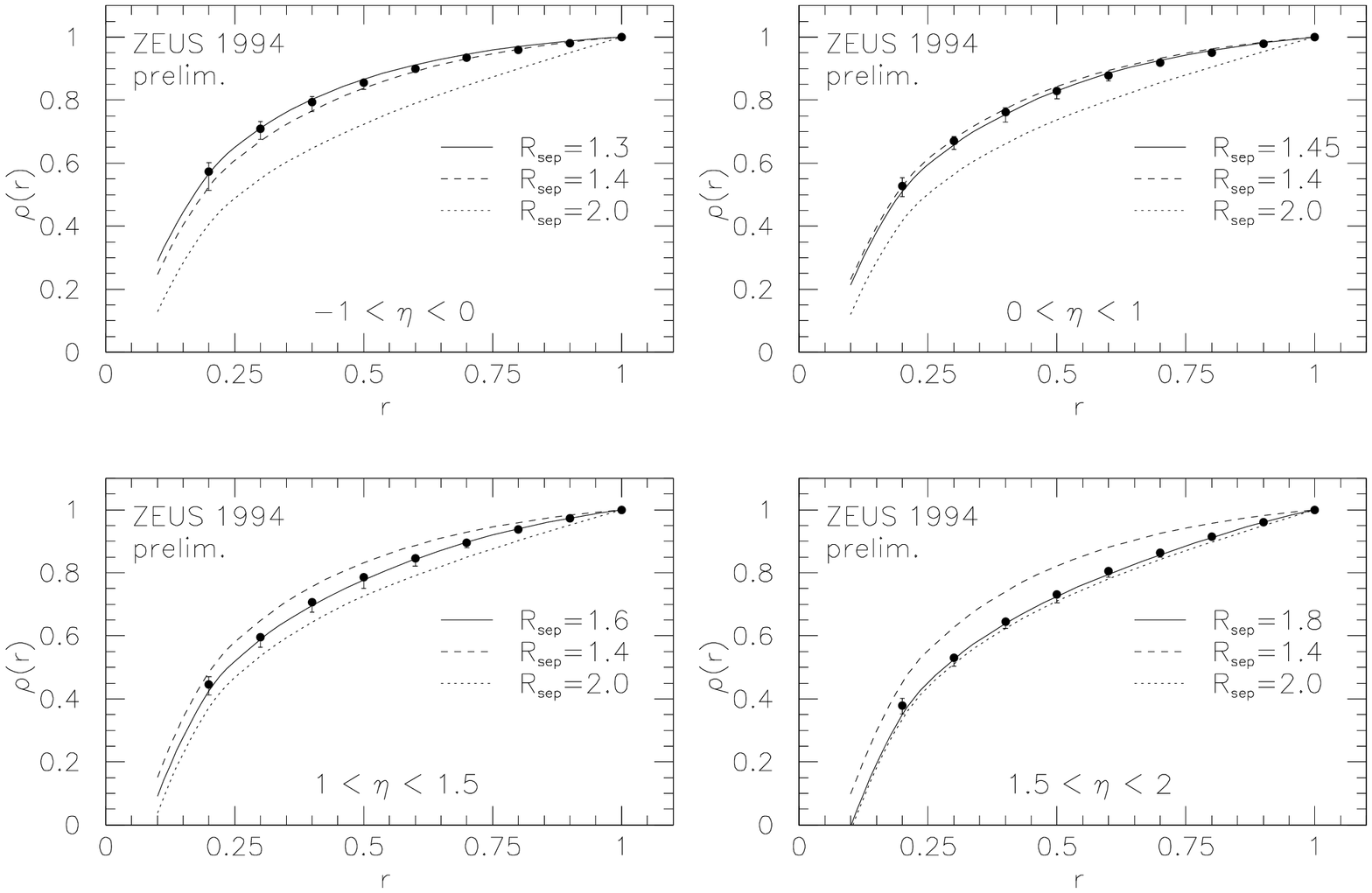,bbllx=520pt,bblly=95pt,bburx=105pt,bbury=710pt,%
           height=14cm,clip=,angle=270}
  \end{picture}}
 \end{center}
 \fcaption{\it Jet shape $\rho(r)$ for
           single-jet photoproduction integrated over $E_T>14$ GeV and four
           different regions of $\eta$. We compare our results using the
           Snowmass convention with $R=1$ and three different values of
           $R_{\rm sep}$ to preliminary 1994 data from ZEUS.}
%
 \begin{center}
  {\unitlength1cm
  \begin{picture}(14,9.5)
   \epsfig{file=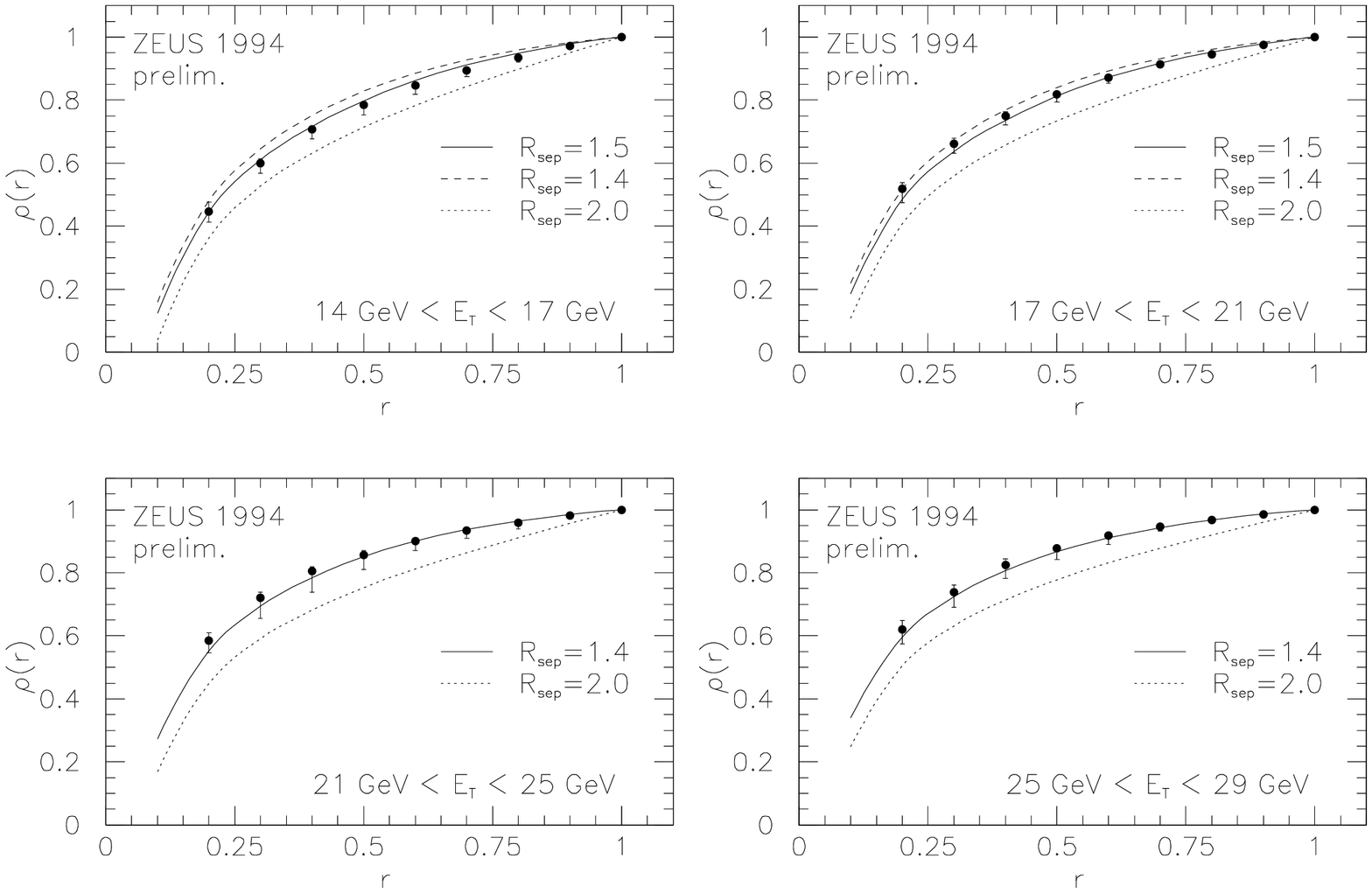,bbllx=520pt,bblly=95pt,bburx=105pt,bbury=710pt,%
           height=14cm,clip=,angle=270}
  \end{picture}}
 \end{center}
 \fcaption{\it Jet shape $\rho(r)$ for
            single-jet photoproduction integrated over $-1<\eta<2$ and four
            different regions of $E_T$. We compare our results using the
            Snowmass convention with $R=1$ and different values of
            $R_{\rm sep}$ to preliminary 1994 data from ZEUS.}
\end{figure}

In a recent study\cite{x15}, we used this $R_{\rm sep}$ parameter to describe
jet shapes $\rho(r)$ as measured by the ZEUS, CDF, and D0 collaborations. The
result is compared to preliminary ZEUS data from 1994\cite{x16} which was
obtained with an iterative cone algorithm and is shown
in Figs.~1 and 2. It depends moderately on the rapidity $\eta$ of the jet
(Fig.~1) being broader in the forward direction and narrows slightly with the
transverse energy $E_T$ of the jet (Fig.~2). Whereas the curves with $R_{\rm
sep}=2$ correspond to no $R_{\rm sep}$ and predict too broad jets, an average
value of $R_{\rm sep}=1.4$ describes the data rather well.

The uncertainties of the Snowmass cone definition are circumvented in the
hadronic implementation of the $k_T$-cluster algorithm\cite{x17} with the
same $E_T$-weighted recombination scheme as for Snowmass, but a different jet
condition. One considers the distance of two particles in $\eta-\phi$ space.
This corresponds to a unique value of $R_{\rm sep}=R$ in theory and
experimentally assigns every hadron to a unique jet.

\section{Real and Virtual Photon Structure}

The determination of the structure of the photon is clearly one of the most
important physics goals in photoproduction. Since the cross section drops
rapidly with the transverse energy, most events at HERA have been observed at
small $E_T$ so far. They are dominantly produced by resolved photons and give
access to the poorly constrained small-$x$ region and the gluon density in the
photon.

However at low $E_T$ a separation of hard and soft physics is experimentally
and theoretically difficult. Hadronization effects, jet definition
uncertainties, energy pedestals around the jets, and multiple interactions
between the remnant jets may play an important role. This can be seen in
Figs.~3 and 4, where we compare our NLO calculation to preliminary 1994 data
from ZEUS, again obtained with the iterative cone algorithm\cite{x18}.
At $E_T > 14$ GeV,
there is a clear excess of data over theory in the forward direction, which
decreases continuously for larger $E_T$. We do not show a comparison with data
obtained with a smaller cone size $R=0.7$ which shows no excess in the forward
region. Obviously the hard jets are better separated from the underlying event
for smaller cones and there is no need for multiple interaction effects.

A second conclusion from Figs.~3 and 4 is that the jet definition uncertainty
is of comparable size to the photon structure function uncertainty. Thus one
either has to rely on the fitted value of $R_{\rm sep} = 1.4$ from jet shapes
or use the $k_T$ cluster algorithm.

In the backward region, where direct and quark initiated processes dominate,
there is fairly good agreement with the data with a slight tendency of GRV to
overestimate the measurement. This can be understood from the bigger quark
contribution of GRV compared to GS96 at large values of $x$.

\begin{figure}[h]
 \begin{center}
  {\unitlength1cm
  \begin{picture}(14,8.5)
   \epsfig{file=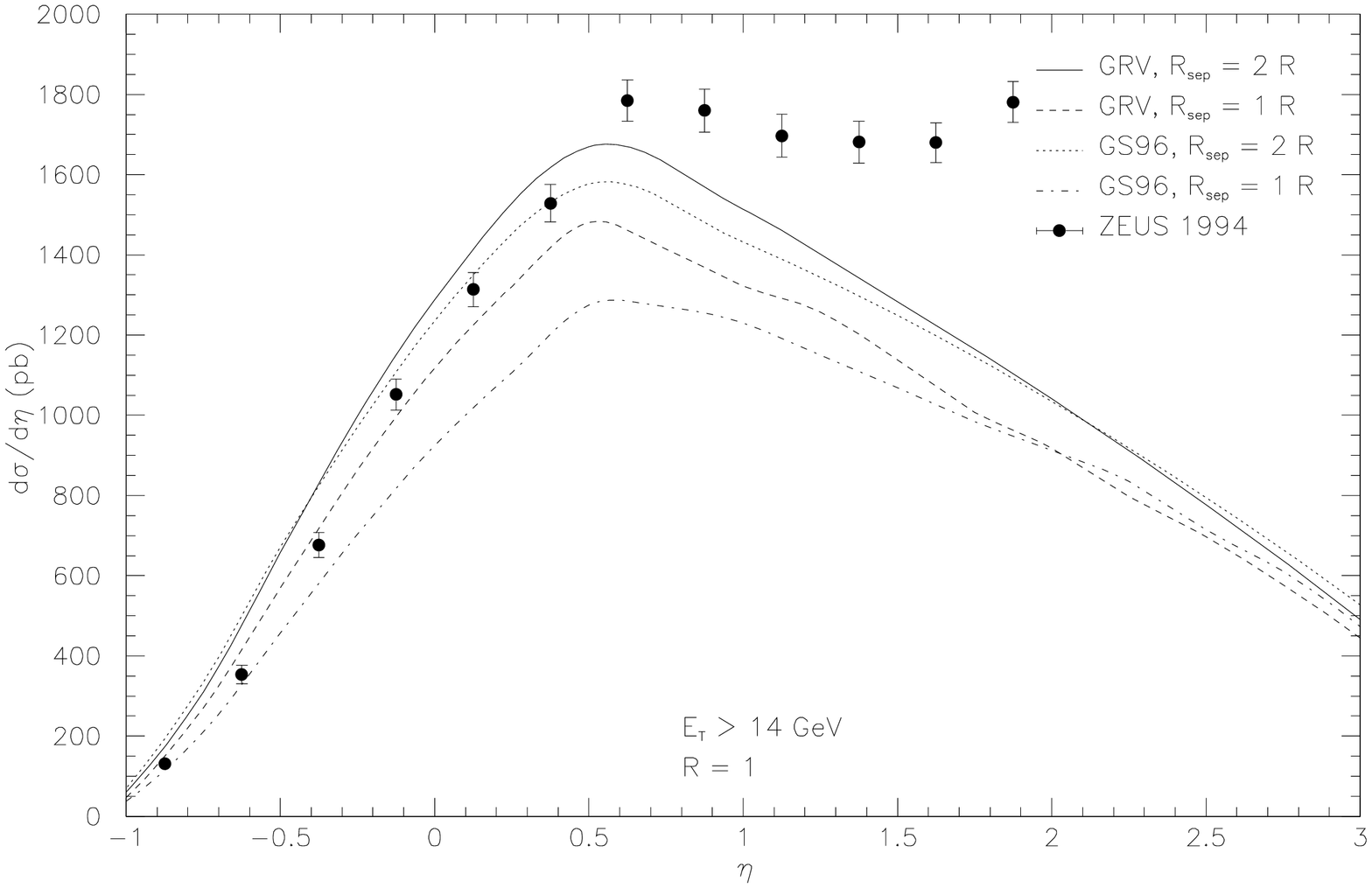,bbllx=520pt,bblly=95pt,bburx=105pt,bbury=710pt,%
           height=14cm,clip=,angle=270}
  \end{picture}}
 \end{center}
 \fcaption{\it Rapidity dependence of the single-jet photoproduction cross
           section integrated above $E_T > 14$ GeV. We compare our NLO
           prediction with GRV and GS96 photon parton densities and the two
           extreme values of $R_{\rm sep}=1R,2R$ to preliminary 1994 data from
           ZEUS.}
\end{figure}
\begin{figure}[h]
 \begin{center}
  {\unitlength1cm
  \begin{picture}(14,9.5)
   \epsfig{file=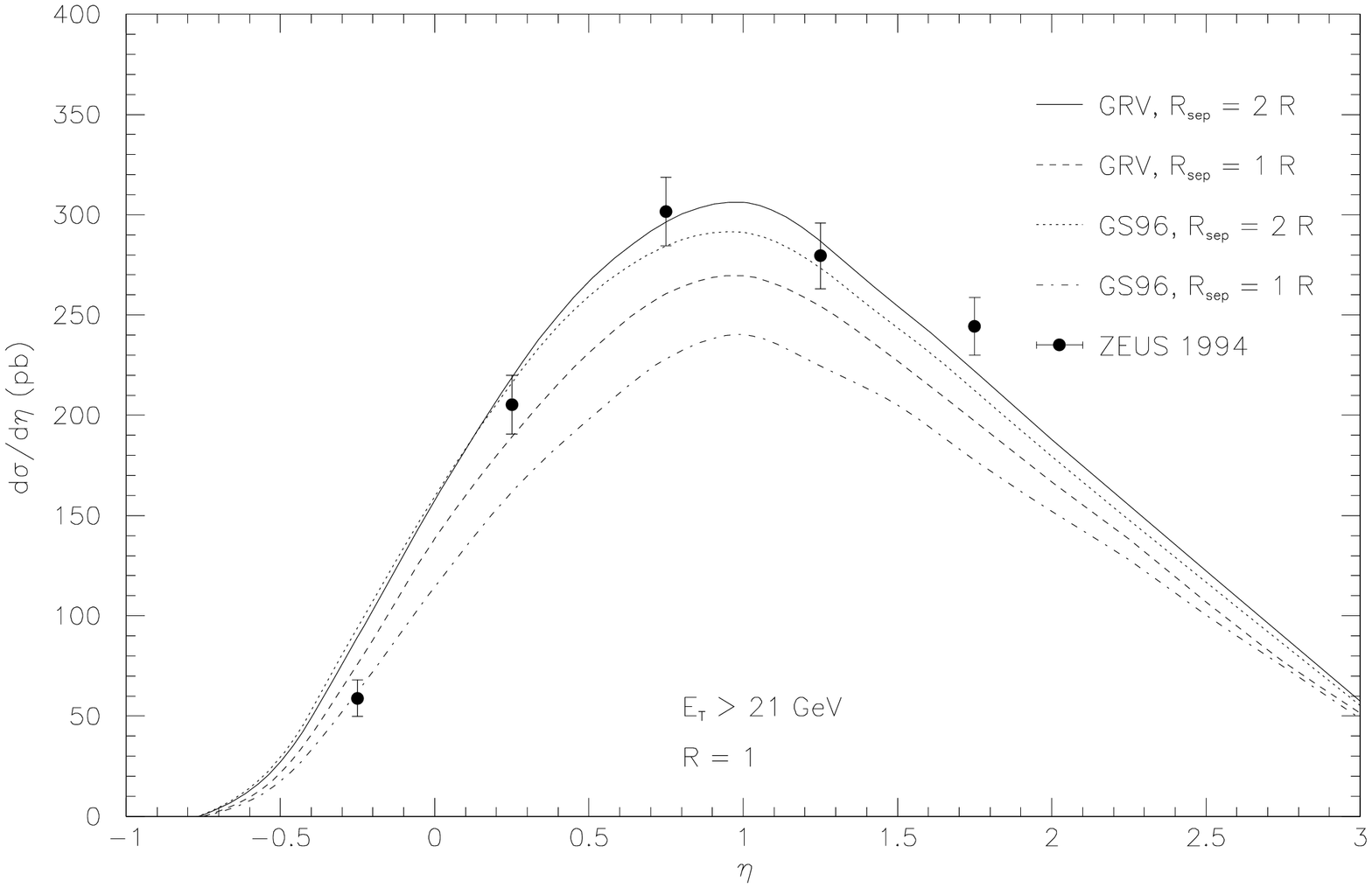,bbllx=520pt,bblly=95pt,bburx=105pt,bbury=710pt,%
           height=14cm,clip=,angle=270}
  \end{picture}}
 \end{center}
 \fcaption{\it Same as Fig.~3 for $E_T > 21$ GeV.}
\end{figure}
\begin{figure}[h]
 \begin{center}
  {\unitlength1cm
  \begin{picture}(14,19)
   \epsfig{file=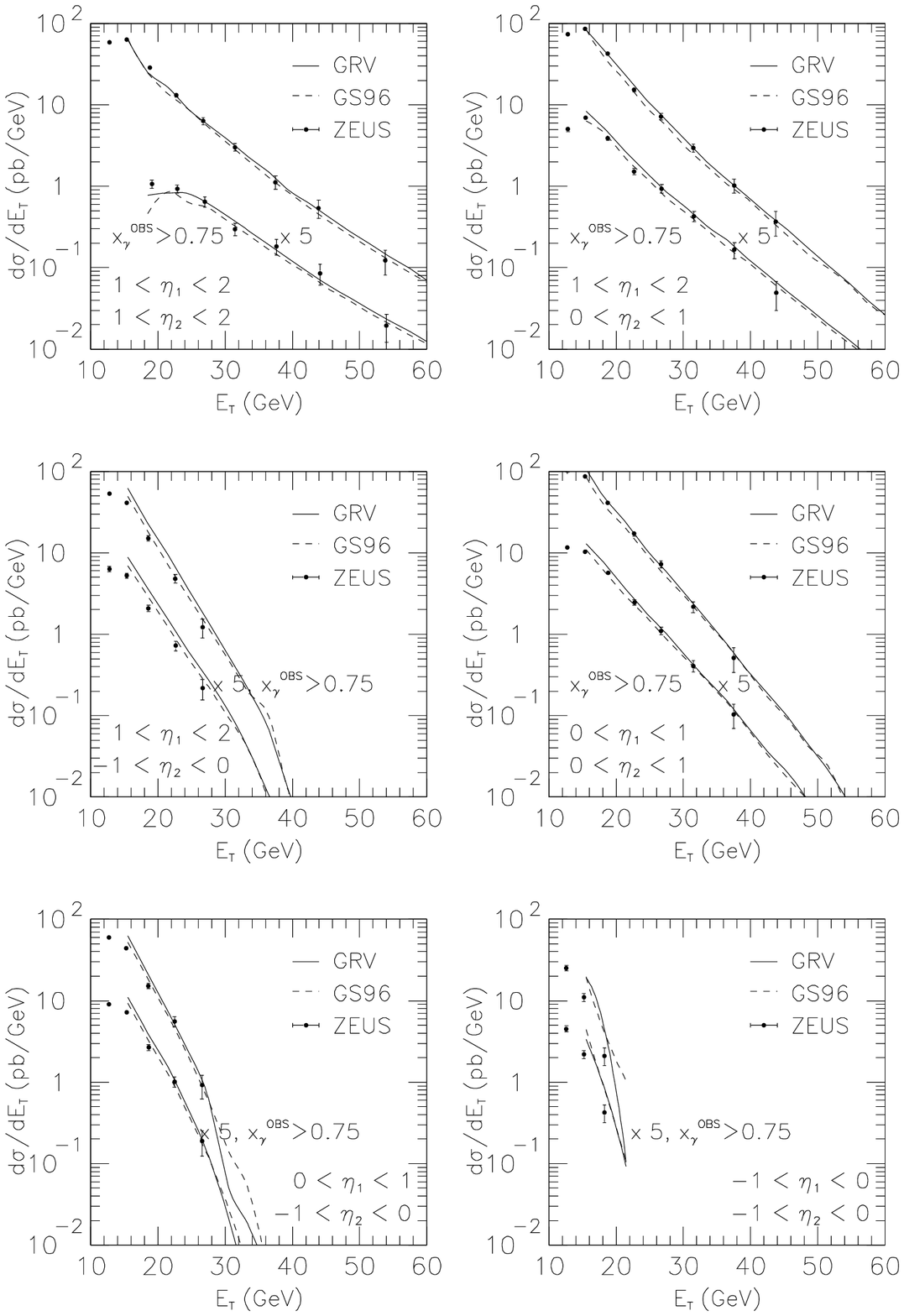,bbllx=70pt,bblly=105pt,bburx=490pt,bbury=715pt,%
           height=19cm,clip=}
  \end{picture}}
 \end{center}
 \fcaption{\it $E_T$ dependence of the symmetrized dijet photoproduction cross
           section integrated over different rapidity bins. We compare our NLO
           prediction with GRV and GS96 photon parton densities and the full
           and upper range of $x_{\gamma}^{\rm OBS}$ to preliminary 1995 data
           from ZEUS.}
\end{figure}
\begin{figure}[h]
 \begin{center}
  {\unitlength1cm
  \begin{picture}(14,19)
   \epsfig{file=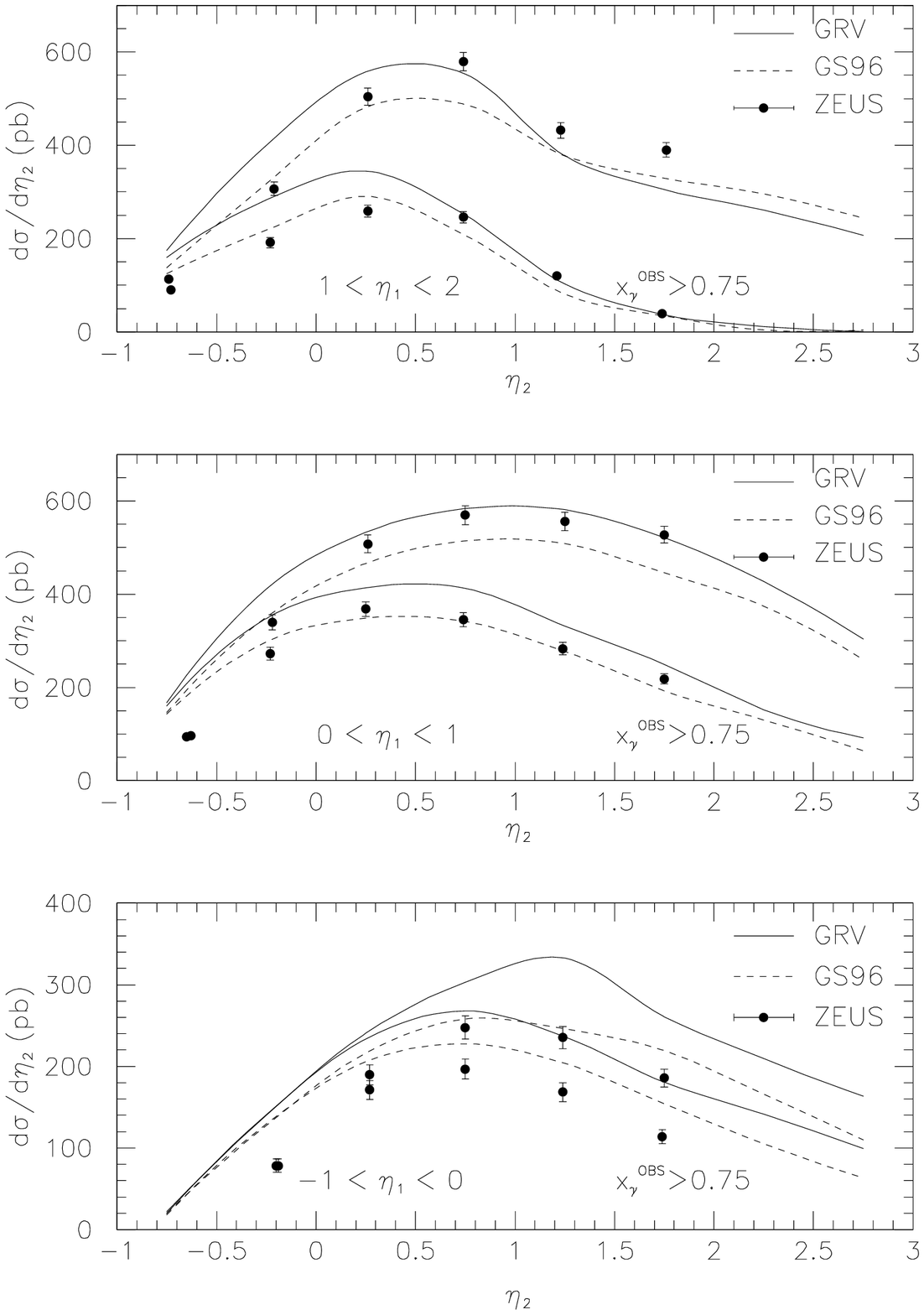,bbllx=60pt,bblly=100pt,bburx=490pt,bbury=715pt,%
           height=19cm,clip=}
  \end{picture}}
 \end{center}
 \fcaption{\it $\eta_2$ dependence of the symmetrized dijet photoproduction
           cross section integrated over $E_T > 14$ GeV and $E_{T_2} > 11$ GeV.
           We compare our NLO prediction with GRV and GS96 photon parton
           densities and the full and upper range of $x_{\gamma}^{\rm OBS}$ to
           preliminary 1995 data from ZEUS.}
\end{figure}
\begin{figure}[h]
 \begin{center}
  {\unitlength1cm
  \begin{picture}(12.5,12.5)
   \put(0,6.75){\epsfig{file=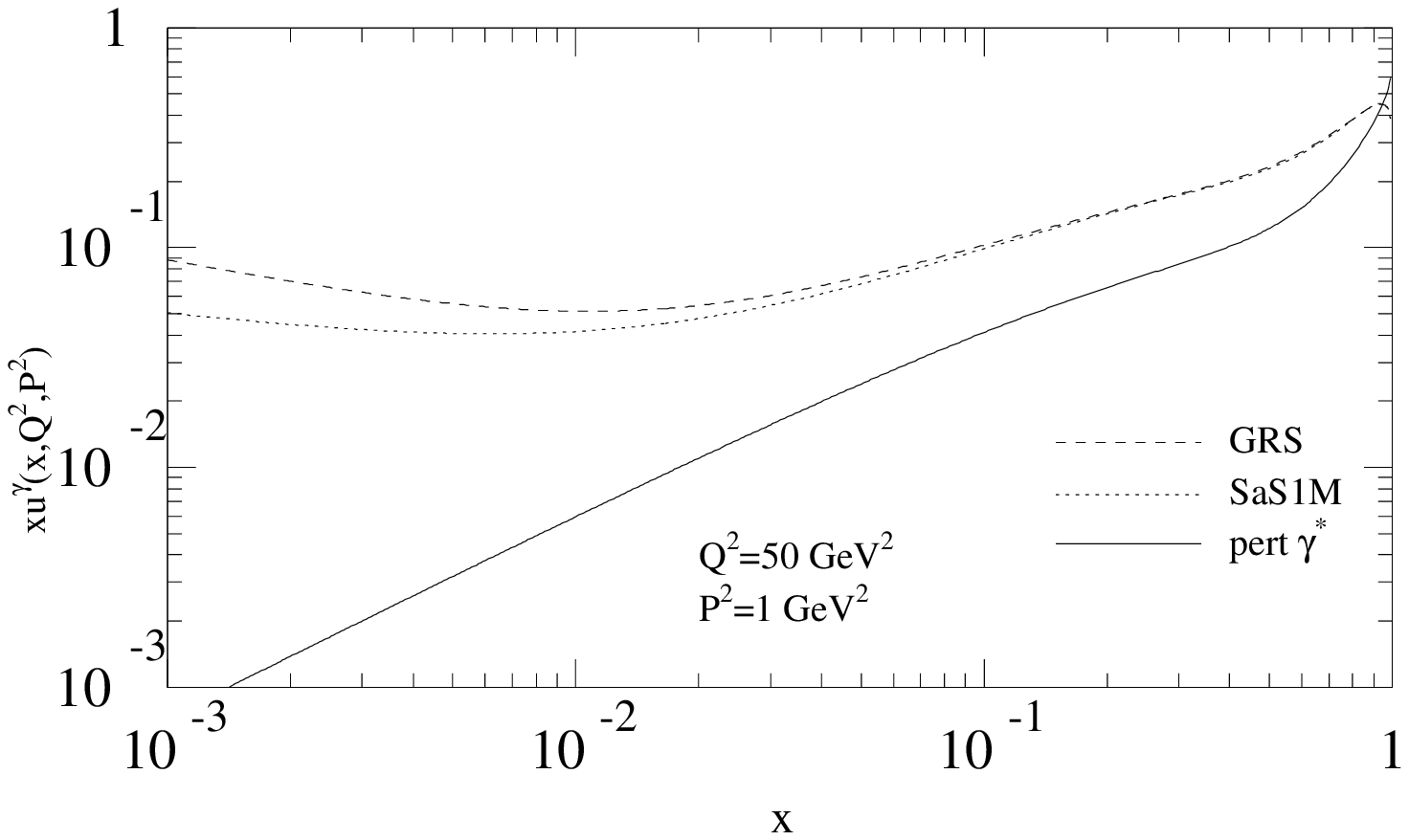,width=12.5cm,clip=}}
   \put(0,0){\epsfig{file=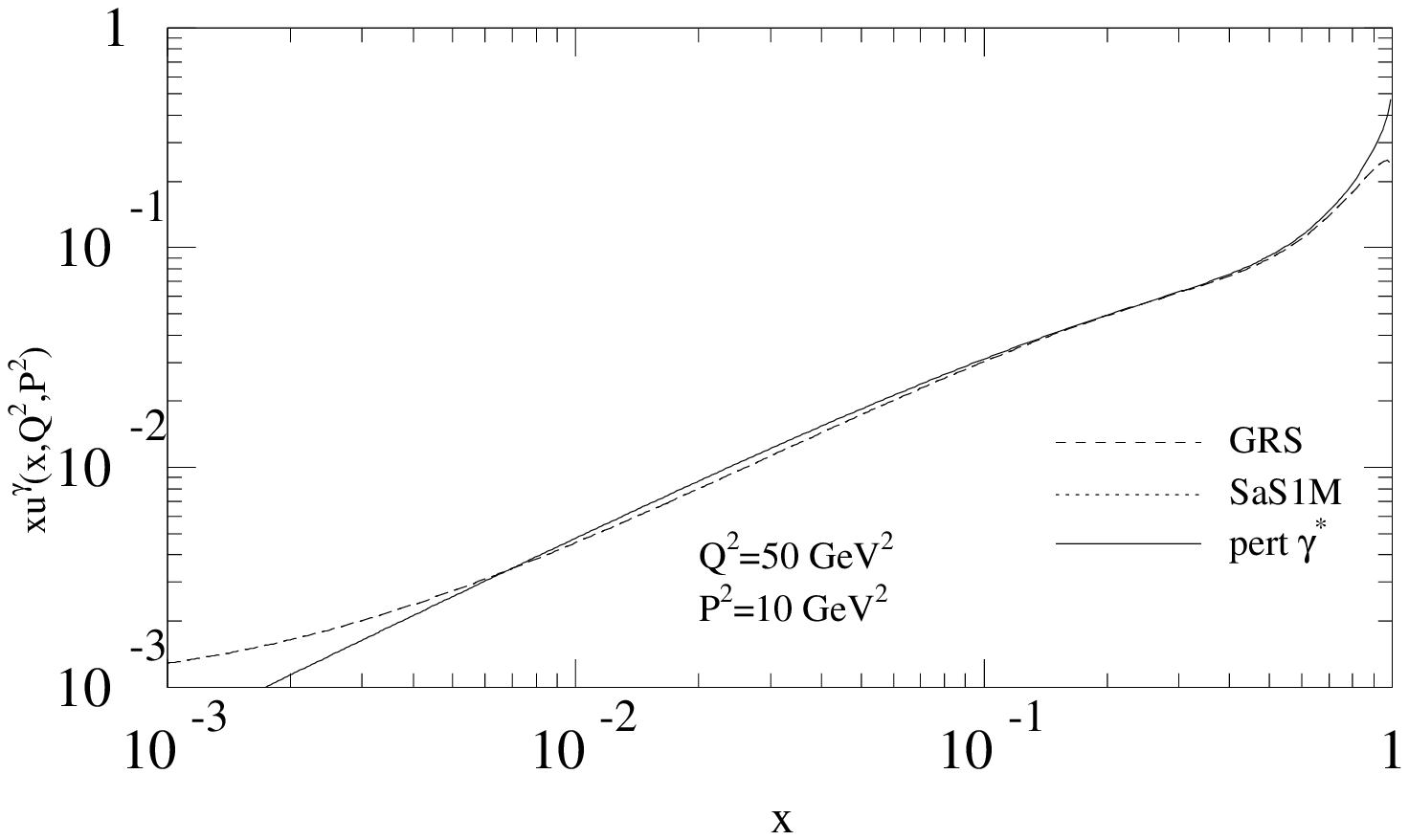,width=12.5cm,clip=}}
  \end{picture}}
 \end{center}
 \fcaption{\it Up-quark distributions in the virtual photon as function of
           $x$. We compare the GRS and SaS1M parametrizations with the
           perturbative result for $P^2 = 1$ and $10$ GeV$^2$ \cite{x24}.}
%
 \begin{center}
  {\unitlength1cm
  \begin{picture}(13,6.5)
   \epsfig{file=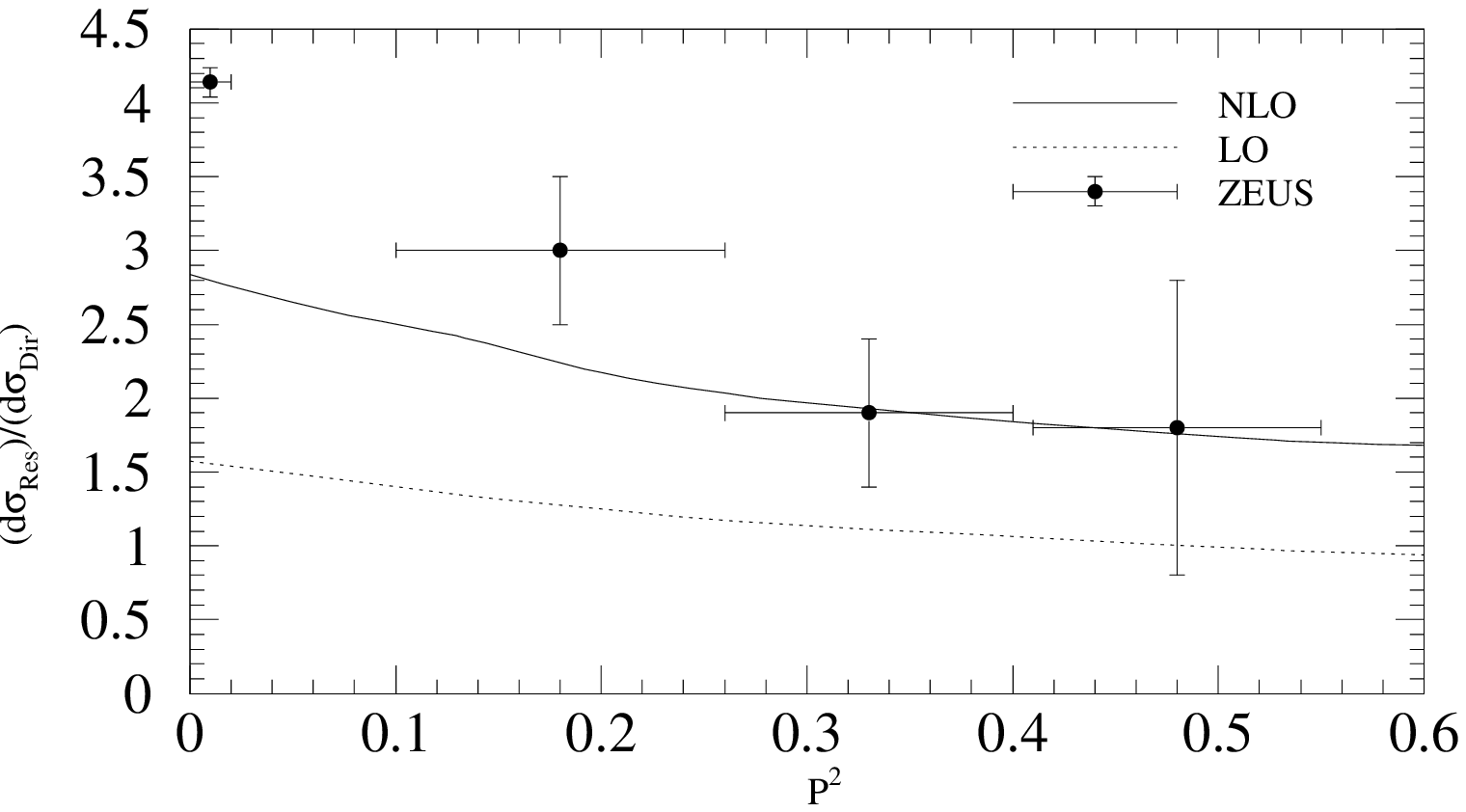,width=12.5cm,clip=}
  \end{picture}}
 \end{center}
 \fcaption{\it Ratio of resolved over direct contributions defined with the
           $x_{\gamma}^{\rm OBS}$ cut at 0.75 for $E_{T_1}, E_{T_2} > 4$ GeV
           using the SaS1M virtual photon structure with four flavors.}
\end{figure}

The kinematics of the partonic subprocess can be better constrained in dijet
cross sections than in single cross sections, e.g. with the help of the
variable
 $x_{\gamma}^{\rm OBS}=(E_{T_1}e^{-\eta_1}+E_{T_2}e^{-\eta_2})/(2yE_e)$.
As long as one integrates over different bins, this variable is infrared-safe
and there is no need to abandon it as put forward by Aurenche et al.\cite{x8}
With the help of this variable, ZEUS separated direct and resolved regions
experimentally and studied the dependence of the cross section on the average
rapidity of the observed jets\cite{x19}. Unfortunately they constrained the
$E_T$ of both jets to the same minimal values. The theoretical predictions are
then not infrared safe and depend on a phenomenological (Klasen and Kramer)
or the technical cut-off (Harris and Owens)\cite{x8}. This analysis employed
the $k_T$ cluster algorithm with $R=1$ and exhibited again the excess of
data over theory for the resolved dominated regions at low $x_{\gamma}^{\rm
OBS}$ and $E_T$.

A new preliminary analysis of 1995 ZEUS data has been presented
recently\cite{x20}. They measured the symmetrized dijet cross section
d$\sigma$/d$E_T$d$\eta_1$d$\eta_2$ with the leading $E_T > 14$ GeV and
$E_{T_2} > 11$ GeV in different rapidity bins and the full and upper region of
$x_{\gamma}^{\rm OBS}$. The result is infrared safe and compared to our NLO
predictions in Fig.~5 as a function of $E_T$ and in Fig.~6 as a function
of $\eta_2$. The general agreement is good for both GRV and GS96, even when
both jets are in the forward region and the complete range in $x_{\gamma}^{\rm
OBS}$ is covered. This may be due to the fairly large cuts on the $E_T$ of both
jets, which suppresses the underlying event. In the backward region, the
calculation lies above the data and GS96 is slightly favored as was the case
for the single-jet cross sections. In addition the systematic errors not
shown here are quite prominent in the backward region. Since the data using
the $k_T$ algorithm were not yet available, we simulated the iterative cone
algorithm with the optimized value of $R_{\rm sep} = 1.4$ for $R=1$. This value
should be lower in the backward region leading to a reduced theoretical
prediction and better agreement with the data.

A new NLO program for virtual photoproduction\cite{x9}
allows us to study the the
transition to photons with virtuality $P^2$. Apart from using the
unintegrated Weizs\"acker-Williams approximation, the main difference
consists in the analytic integration of the virtual photon initial state
singularity. This singularity is then factorized and leads to a
scheme- and $P^2$-dependent finite contribution which reduces to the real
expression as $P^2\rightarrow 0$. How this leading-logarithmic singularity
compares to the two existing up-quark parametrizations in the virtual
photon\cite{x21} is shown in Fig.~7. At low $P^2 = 1$ GeV$^2$ the evolution
up to 50 GeV$^2$ produces many more up-quarks than just the single pair
predicted perturbatively. At large $P^2 = 10$ GeV$^2$, GRS and SaS1M agree
with the perturbative box-diagram\cite{x24}.

ZEUS have also published $P^2$ dependent data on the ratio of resolved over
direct contributions disentangled with the $x_{\gamma}^{\rm OBS}$ cut at
0.75\cite{x22}. As can be seen in Fig.~8 the NLO effects are large at
$E_{T_1}, E_{T_2} > 4$ GeV. The lowest $P^2$ point for real photoproduction
has much better statistics than the virtual photoproduction data and lies
above the theory. This is to be expected since we are again in the soft region
where remnant-remnant interactions are assumed to be important. At larger
$P^2$, theory and data agree well in shape and normalization. We use the SaS1M
LO parametrization, since NLO parton densities are not available in
parametrized form and GRS does not contain charm.

\section{Proton Structure at large $x$}
\begin{figure}[p]
 \begin{center}
  {\unitlength1cm
  \begin{picture}(14,8.5)
   \epsfig{file=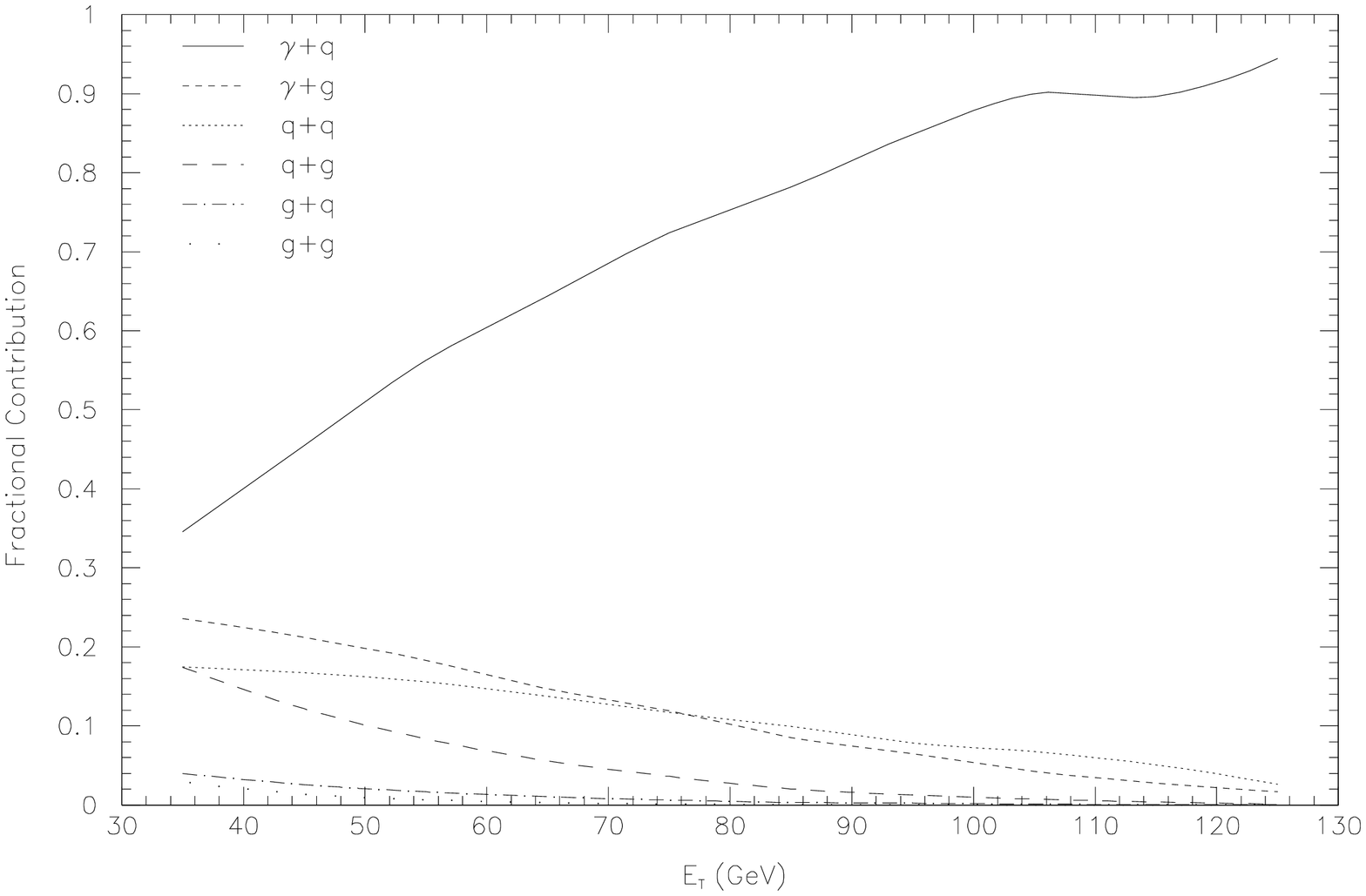,bbllx=520pt,bblly=95pt,bburx=105pt,bbury=710pt,%
           height=14cm,clip=,angle=270}
  \end{picture}}
 \end{center}
 \fcaption{\it Fractional Contribution of direct and resolved partonic
           subprocesses as function of $E_T$ in the single-jet photoproduction
           cross section.}
%
 \begin{center}
  {\unitlength1cm
  \begin{picture}(14,9.5)
   \epsfig{file=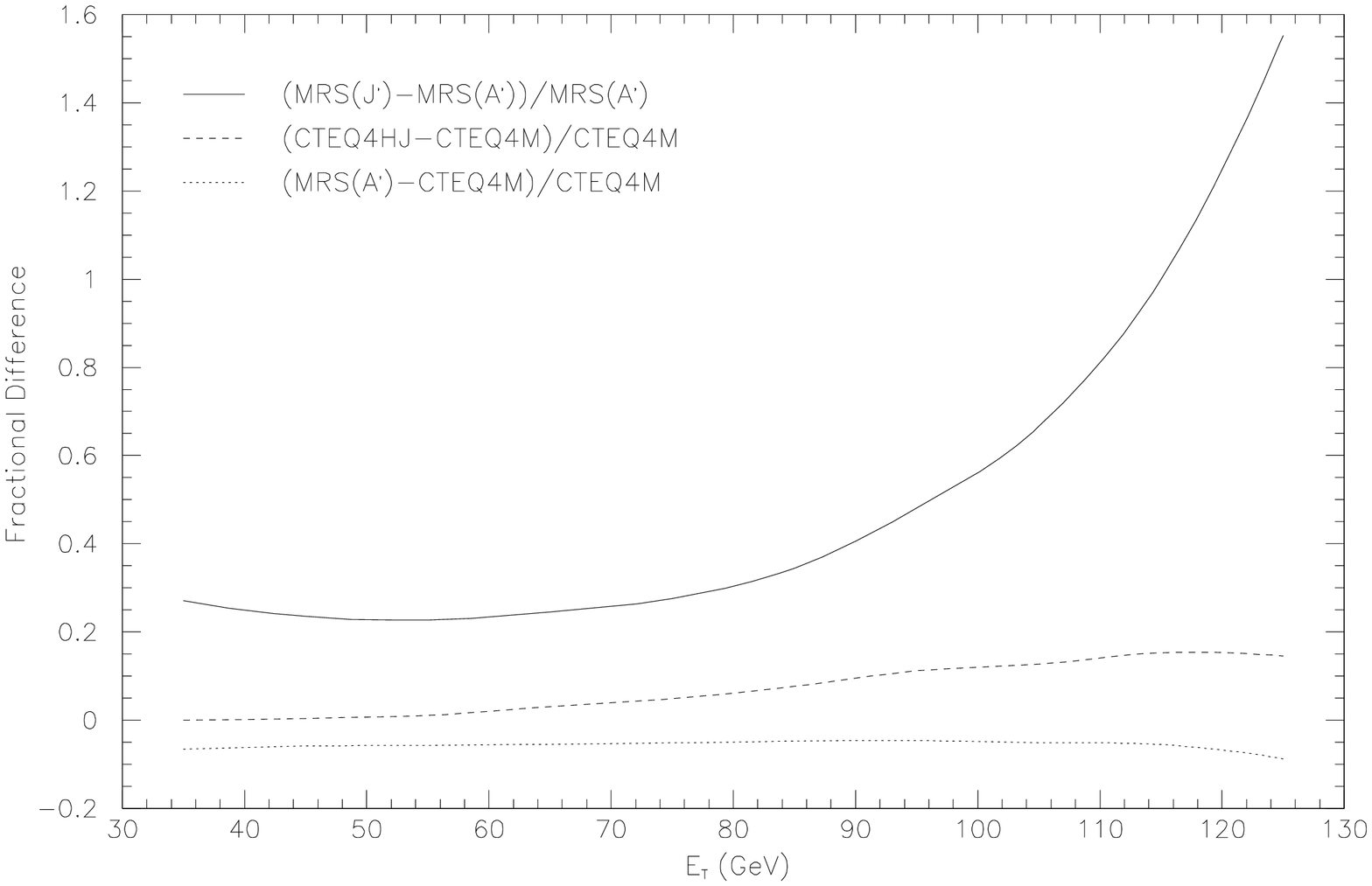,bbllx=520pt,bblly=95pt,bburx=105pt,bbury=710pt,%
           height=14cm,clip=,angle=270}
  \end{picture}}
 \end{center}
 \fcaption{\it Fractional difference of different proton structure functions
           as function of $E_T$ in single-jet photoproduction compared to
           standard parametrizations.}
\end{figure}
The excess of high-$Q^2$ events in DIS at HERA have triggered a lot of
speculation about new physics such as leptoquarks, $R$-parity violating
supersymmetry, or contact interactions. This kinematic regime can and
should also be tested in photoproduction at large $E_T$, e.g.~by measurements
of d$\sigma$/d$E_T$, d$\sigma$/d$E_{\gamma}$, or d$\sigma$/d$M_{JJ}$. An
unambiguous determination of new physics requires a precise knowledge of the
Standard Model background. Due to phase space limitations, NLO corrections
become increasingly important at the boundary of phase space at large $E_T$,
whereas the uncertainty from the photon structure function becomes negligible.
This can be extracted from Fig.~9, where the fractional contribution of
direct photon-quark scattering accounts for more than 90\% of the total
single-jet cross section above $E_T = 100$ GeV. Therefore a precise knowledge
of the quark distribution in the proton is required for $x > 0.4$.

In Fig.~10, we demonstrate the increase in the cross section that can be
obtained with the larger quark densities of MRS(J') which were designed to
describe the excess of CDF $p\overline{p}$ data at large $E_T$\cite{x23}. A
factor of 1.5 seems possible here. Larger gluon distributions as proposed by
CTEQ4(HJ)\cite{x11} only give a rise of 15\%. It should be mentioned that
MRS(J') fail to fit the low energy BCDMS data, whereas CTEQ4(HJ) are still
consistent. Finally, a study of the scheme dependence might indicate further
uncertainties here or in the gluon density.

\section{Conclusions}
A wealth of new and precise data coming from H1 and ZEUS in photoproduction
has triggered increased theoretical interest. Three independent NLO
calculations for real dijet production and the first NLO program for virtual
photoproduction are now available. The interpretation of jets is improved
with smaller cone sizes and the $k_T$ cluster algorithm. This will allow for
first stringent tests of the real photon parton density, especially of the
gluon. More data and a parametrization of NLO parton densities are needed in
virtual photoproduction. The access to large transverse momenta offers
interesting studies of the proton structure at large $x$.

\section{Acknowledgements}
The author thanks the organizers of the Ringberg workshop for the kind
invitation, G.~Kramer and B.~P\"otter for their collaboration, and C.~Glasman
and Y.~Yamazaki for making the preliminary ZEUS data available to him.

\end{document}